# Rotating frame spin dynamics of a Nitrogen-Vacancy center in a diamond nanocrystal

Abdelghani Laraoui[1] and Carlos A. Meriles[1,*]

[1]Department of Physics, City College of New York – CUNY, New York, NY 10031, USA

**Abstract**

We investigate the spin dynamics of a Nitrogen-Vacancy (NV) center contained in an individual diamond nanocrystal in the presence of continuous microwave excitation. Upon periodic reversal of the microwave phase, we observe a train of 'Solomon echoes' that effectively extends the system coherence lifetime to reach several tens of microseconds, depending on the microwave power and phase inversion rate. Starting from a model where the NV center interacts with a bath of paramagnetic defects on the nanocrystal surface, we use average Hamiltonian theory to compute the signal envelope from its amplitude at the echo maxima. Comparison between the effective Rabi and Solomon propagators shows that the observed response can be understood as a form of higher-order decoupling from the spin bath.

---

[*] To whom correspondence should be addressed. E-mail: cmeriles@sci.ccny.cuny.edu

The spin angular momentum associated with a NV center—a substitutional Nitrogen impurity adjacent to a vacancy in the diamond lattice—is emerging as a promising platform for nanoscale magnetic sensing and imaging[1-3]. Among the many properties that make this system attractive are its outstanding photostability and long spin lifetimes at room temperature. Also key is the ability to optically initialize and readout the spin state, which, in turn, can be readily manipulated via the application of microwave pulses. Of special interest in the context of nanoscale sensing are the NV centers confined within diamond nanocrystals, already used as local probes when embedded in heterogeneous environments such as biological systems[4-6]. Indeed, because diamond is non-cytoxic and relatively easy to functionalize, NV-containing nanocrystals have proven to serve as efficient fluorescent biomarkers for single particle tracking[4]. More importantly, recent work within living cells has shown that the NV spin can be monitored individually and in real time, and that the cell environment is not necessarily detrimental to spin coherence[5,6].

Different schemes have been proposed to probe the nanocrystal dynamics (and through it, the surrounding medium), one possibility being the application of long microwave pulses resonant with one NV spin transition[7]. In this modality the particle position and orientation can be extracted from the frequency of the resulting Rabi oscillations, itself a function of the distance and crystal-field alignment relative to the microwave source. As is the case in magnetic resonance microscopy, the spatial resolution of this imaging strategy is not dependent on diffraction but rather dictated by the microwave field gradient and the effective line width of the chosen spin transition. Unfortunately, and unlike bulk crystal, NV spins in nano-sized diamond are prone to exhibit relatively short Rabi lifetimes, typically not exceeding a few hundred nanoseconds[8]. Here we investigate the spin dynamics of a single NV center within an individual ~70 nm diamond crystal, and show that significantly longer lifetimes can be attained by periodically alternating the microwave phase. We observe a pattern of 'Solomon' (or 'rotary') echoes which, as we demonstrate analytically, results from high-order truncation of the non-secular contributions to the effective rotating-field Hamiltonian. In agreement with our formalism, we find that the effective lifetime grows with the microwave power and phase-change rate.

For the experiments reported herein we use commercial nanodiamond (Microdiamant) with average diameter of 70±20 nm. We follow a simple sample preparation protocol whereby nanocrystals are dissolved in ethanol and drop-cast on a clean glass substrate; the AFM image of a typical diamond particle affixed to the glass surface after solvent evaporation is shown in Fig. 1a. To address the NV centers within a given nanocrystal we use a homemade confocal microscope with laser excitation at 532 nm[9,10]. For particles of this size the average number of NV centers per crystal is low[11]—approximately, 0.2—implying that most fluorescing nanocrystals contain only one color center. The NV has spin angular momentum $S=1$ and a triplet ground state with splitting $\Delta=2.87$ GHz, the so-called 'crystal field' pointing along the <111> crystallographic direction. In our experiments we apply a mild (6.5 mT) magnetic field aligned with the crystal field, which breaks the degeneracy between the $|m_S=\pm1\rangle$ levels. To selectively address one of the two possible transitions (e.g., the $|m_S=0\rangle \rightarrow |m_S=1\rangle$ transition), we use the microwave field produced by a 25-μm-diameter wire overlaid on the glass substrate and adjacent to the diamond particles. The optically-detected magnetic resonance spectrum from an individual diamond nanocrystal as determined from a cw frequency sweep is shown in Fig. 1b.

Undergoing several cycles to attain sufficient statistics, the typical time-resolved magnetic resonance experiment (Fig. 2a) starts with a short (1 µs) light pulse to initialize the NV state to $|m_S = 0\rangle$[12]. After microwave manipulation, we determine the population difference between the $|m_S = 0\rangle$ and $|m_S = 1\rangle$ states from the fluorescence of a readout pulse (0.5 µs), which is then followed by a signal reference pulse (1 µs); the maximum fluorescence contrast between the two spin states is 30 %[12]. Figs. 2b-d show the normalized optical signal from an individual nanocrystal for three representative magnetic resonance sequences, namely the Ramsey, Hahn-echo, and Rabi protocols, exhibiting coherence lifetime of order 0.2, 10 and 0.7 µs respectively. In agreement with prior findings[11], comparison with analog experiments in bulk diamond[13] shows that the observed signals are considerably shorter-lived (approximately 20 to 50 fold in all three cases). The latter was found to be characteristic of all the nanocrystals we tested and is likely caused by the various paramagnetic defects forming at the particle surface[14]. In spite of this difference, we find that the intercalation of an inversion pulse during free evolution still succeeds in extending the Ramsey lifetime (see Figs. 2b and 2c), making us wonder whether a 'Hahn-echo-like' strategy may be applicable to the Rabi sequence as well[13,15,16].

Fig. 3 shows a modified Rabi protocol where the phase of the variable-duration microwave pulse is inverted at fixed, predetermined times and according to a periodic pattern. Already introduced by Solomon in the early days of magnetic resonance[17], this scheme has not yet been applied to single NVs, presumably because the radio-frequency field inhomogeneities it typically aims to compensate for, are absent when probing individual defects. Fig. 3b shows, nonetheless, that the Solomon protocol results in a very singular response exhibiting well-defined echoes at the midpoint between consecutive phase inversions and a slower overall decay rate.

To interpret these observations we start by writing the system Hamiltonian in the presence of static and microwave magnetic fields—respectively $B_0 \hat{z}$ and $B_1 \hat{x}$—as

$$H = \Delta S_z^2 + \Omega_0 S_z + 2\Omega_1 S_x \cos\left(\Omega_{mw} t + \frac{\pi}{2}(1 - f(t))\right) + \sum_{\substack{j,=1..N \\ \alpha=x,y,z}} S_z A_{z\alpha}^{(j)} I_\alpha^{(j)} + \sum_j \Omega_L^{(j)} I_z^{(j)}, \quad (1)$$

where $\Delta$ denotes the crystal field amplitude, and $\Omega_0 \equiv \gamma_{NV} B_0$ with $\gamma_{NV}$ the NV gyromagnetic ratio. We describe the microwave field via the frequency $\Omega_{mw}$, the field amplitude $\Omega_1 \equiv \gamma_{NV} B_1$, and the phase-governing, piecewise-constant function $f(t)$ switching instantaneously between -1 and 1. The NV center interacts with a 'bath' of surrounding, not-necessarily-identical spins $I^{(j)} = 1/2$ coupled to the NV center via the coupling matrices $\mathbf{A}^{(j)}$. The last term expresses the interaction of these spins with the external field $B_0$; for simplicity, we will ignore the dipolar interactions between bath spins. To eliminate the explicit time dependence in $H$, it is customary to transform to a rotating frame via the unitary operator $U_r = \exp(-i\Omega_{mw} t S_z)$. Ignoring rapidly fluctuating terms and assuming $\Omega_L^{(j)} \gg A_{z\alpha}^{(j)}$ (a good approximation for electronic spins with Landé factor near 2 at 6.5 mT), we write the rotating frame Hamiltonian as

$$H_r = \Delta S_z^2 + (\Omega_0 - \Omega_{mw}) S_z + f(t)\Omega_1 S_x + \sum_{j,=1..N} S_z A_{zz}^{(j)} I_z^{(j)} + \sum_j \Omega_L^{(j)} I_z^{(j)}. \quad (2)$$

In the above representation, the observed signal can be calculated from the formula

$$O(t) = Tr\{V^\dagger \rho_0 V \sigma_z\}, \quad (3)$$

where $V = \exp(-iH_r t)$, $\sigma_z \equiv |0\rangle\langle 0| - |1\rangle\langle 1|$ is the Pauli operator in the $\{m_S = 0,1\}$ manifold, and we model the density matrix at time $t=0$ as $\rho_0 = (1/2^N)|0\rangle\langle 0| \otimes \mathbf{1}_I$, with $\mathbf{1}_I$ denoting the identity operator in the spin bath space.

In our experiment the microwave field is tuned to the $|0\rangle \to |1\rangle$ transition, i.e. $\Omega_{mw} = \Delta + \Omega_0$, making it convenient to rewrite Eq. (2) as the sum of three commuting terms, $H_r = H_A + H_B + H_C$. The first (and more important) term effectively describes the NV dynamics as a two level system and can be written as

$$H_A = f(t)\Omega_1 \sigma_x - \frac{1}{2}\sum_j A_{zz}^{(j)} I_z^{(j)} \sigma_z, \qquad (4)$$

where we introduced the Pauli operator $\sigma_x \equiv |0\rangle\langle 1| + |1\rangle\langle 0|$. On the other hand, and defining $\mathbf{1}_\sigma \equiv |0\rangle\langle 0| + |1\rangle\langle 1|$, we find that the second contribution is given by $H_B = \sum_j \Omega_L^{(j)} I_z^{(j)} + (1/2)\sum_j A_{zz}^{(j)} I_z^{(j)} \mathbf{1}_\sigma$, which, from Eq. (3), has no effect on the system evolution. Finally, the last term is found to act only on the $\{m_s = -1\}$ manifold, and in the limit $|\Delta - \Omega_0| \gg \Omega_1$ it takes the approximate form $H_C = \left(\Delta - \Omega_0 + \sum_j A_{zz}^{(j)} I_z^{(j)}\right)|-1\rangle\langle -1|$, with no role on the NV time dynamics. Taking into consideration all of the above, we rewrite Eq. (3) as

$$O(t) = Tr\{U\rho_0 U^\dagger \sigma_z\} \qquad (5)$$

with $U(t) = \exp(-iH_A t)$. It is worth noting that the two contributions to $H_A$—which we will respectively refer to as $H_{Ax}$ and $H_{Az}$—do not commute. Therefore, in the limit of strong microwave irradiation, the NV evolution is governed predominantly by the Rabi field ($H_{Ax}$), as the coupling term ($H_{Az}$) can be safely truncated.

Unfortunately, and unlike experiments in bulk diamond, the above limit is difficult to attain under the present conditions: As inferred from the Ramsey decay in Fig. 2b ($T_2^* \sim 200$ ns) the dipolar field produced by surrounding electronic spins at the NV site is of order 0.2 mT, a value comparable to the strongest possible microwave fields[18] (in our experiments reaching up to 0.7 mT). Therefore, and to more accurately describe the system spin dynamics, we make use of average Hamiltonian theory to cast the evolution operator in the form[19]

$$U(t) = U_x(t) U_z(t), \qquad (6)$$

with $U_x(t) = T \exp\left(-i \int_0^t dt' H_{Ax}(t')\right)$ and $U_z(t) = T \exp\left(-i \int_0^t dt' \hat{H}_{Az}(t')\right)$; $T$ is the Dyson time-ordering operator and $\hat{H}_{Az}(t') = U_x^{-1} H_{Az} U_x$. Because the Solomon sequence follows a cyclic pattern with period $\tau$ (i.e., $U_x(\tau) = \mathbf{1}$), we can calculate the propagator after $m$ periods as $U(t_m \equiv m\tau) = (U_z(\tau))^m$. The latter can be computed perturbatively using the Magnus expansion

$$U_z(\tau) = \exp(-i\tau \overline{H}_{Az}) = \exp\left(-i\tau\left(\overline{H}_{Az}^{(0)} + \overline{H}_{Az}^{(1)} + \overline{H}_{Az}^{(2)} + \cdots\right)\right), \qquad (7)$$

where $\overline{H}_{Az}^{(0)} = \frac{1}{\tau} \int_0^\tau dt\, \hat{H}_{Az}(t)$, $\overline{H}_{Az}^{(1)} = -\frac{i}{2\tau} \int_0^\tau dt \int_0^t dt' \left[\hat{H}_{Az}(t), \hat{H}_{Az}(t')\right]$, etc (see also Ref. (19)).

We can now more thoroughly understand the longer coherence times observed in the NV response to the Solomon sequence (where the phase cycle is defined by the function $f(t) = 1$ for $0 < t < \tau/4$ or $3\tau/4 < t < \tau$ and $f(t) = -1$ for $\tau/2 < t < 3\tau/4$, repeating periodically thereafter, see Fig. 3). For example, choosing for simplicity the microwave power so that $\Omega_1 \tau = 4n\pi$ with $n$ integer, we find after some algebra that the first two terms in the expansion of the average 'Solomon' Hamiltonian $^{Sl}\overline{H}_{Az}$ cancel, i.e., $^{Sl}\overline{H}_{Az}^{(0)} = {}^{Sl}\overline{H}_{Az}^{(1)} = 0$. On the other hand, a direct calculation of the next term in the series yields $^{Sl}\overline{H}_{Az}^{(2)} \propto (\sigma_z/\Omega_1^2)\sum_{j,k,l} A_{zz}^{(j)} A_{zz}^{(k)} A_{zz}^{(l)} I_z^{(j)} I_z^{(k)} I_z^{(l)}$, a contribution that commutes with $\rho_0$ and $\sigma_z$, and that therefore does not lead per se to decoherence (see Eqs. (5)-(7)). Hence, we conclude that the effective Solomon Hamiltonian is given by $^{Sl}\overline{H}_{Az} \sim {}^{Sl}\overline{H}_{Az}^{(2)} + {}^{Sl}\overline{H}_{Az}^{(3)} + {}^{Sl}\overline{H}_{Az}^{(4)} + \cdots$, allowing us to model the Solomon envelope at sufficiently short times $t_m = m\tau$ in the form

$$O_{Sl}(t_m) \sim 1 - \frac{1}{2}\left(\frac{t_m A^{n+1}}{q\Omega_1^n}\right)^2, \qquad (8)$$

Eq. (8) must be understood as a crude estimate in which $q$ is a constant (of order $10^1$-$10^2$), $A \equiv \sqrt{\sum_j \left(A_{zz}^{(j)}\right)^2}$ is a parameter characterizing the coupling of the NV center with the surrounding spin bath, and $n$ equals 3 (or 4, or more depending on the next non-commuting term in the average Hamiltonian expansion).

It is worth comparing Eq. (8) with the corresponding evolution in the Rabi field (where $f(t) = 1$, Fig. 2d). In this case, and for times $t_l$ so that $\Omega_1 t_l = 2l\pi$ with $l$ integer, we use Eqs. (5) to (7) to write the Rabi signal as $O_{Rb}(t_l) = Tr\{U_x U_z \rho_0 U_z^\dagger U_x^\dagger \sigma_z\}$ $\approx Tr\{e^{-it_l {}^{Rb}\overline{H}_{Az}^{(1)}} \rho_0 e^{it_l {}^{Rb}\overline{H}_{Az}^{(1)}} \sigma_z\}$. Assuming $t_l$ sufficiently small, we find

$$O_{Rb}(t_l) \sim 1 - \frac{1}{2}\left(\frac{t_l A^2}{32\Omega_1}\right)^2 \qquad (9)$$

The predicted quadratic time dependence is in agreement with our observations at high microwave power (insert in Fig. 4a), precisely the regime where the average Hamiltonian expansion (Eq. (7)) is expected to converge. Something similar can be said of the Solomon envelope, though in this latter case the transition from exponential to gaussian-like shape is less clearly defined. From the comparison of Eqs. (8) and (9) we conclude that the Rabi envelope is expected to decay considerably faster than its Solomon analog, i.e. $T_{2Rb}/T_{2Sl} \sim (A/\Omega_1)^{2(n-1)} << 1$ with $n$ equal 3 (or greater).

The Solomon response for different durations of the microwave phase cycle $\tau$ is illustrated in Fig. 4b for a fixed excitation power (corresponding to a microwave field $\Omega_1$ of 0.17 mT). We find that the decoherence rate progressively slows down for shorter phase cycles when $\tau \leq 4$ µs, a tendency abruptly reversed at shorter times. While a detailed analysis exceeds the scope of this manuscript, we mention that couplings between bath spins (not considered herein, see Eq. (1)) lead to echo-affecting changes in the bath configuration during the intervals immediately preceding and following phase inversion. Naturally, the influence of these changes on the NV evolution can be mitigated with a faster echo formation rate. In the laboratory frame this is equivalent to replacing a single inversion pulse with a Carr-Purcell train, precisely the strategy exploited recently to extend the NV coherence lifetime in bulk diamond[20,21].

The observed shortening of the signal lifetime at the fastest repetition rate ($\tau = 1.6$ µs) is not yet fully understood, but we hypothesize an imperfect canceling of the lower-order terms in the average Hamiltonian expansion when the number of Rabi beatings per period is low. For example, an exact calculation of the first-order term for the Solomon Hamiltonian yields

$$^{Sl}\overline{H}_{Az}^{(1)} = -2\sigma_x \frac{\sin(\Omega_1\tau/4)\sin^2(\Omega_1\tau/8)}{\Omega_1^2\tau} \sum_{j,k} A_{zz}^{(j)} A_{zz}^{(k)} I_z^{(j)} I_z^{(k)}, \qquad (10)$$

which vanishes only when $\Omega_1\tau = 4m\pi$ but can be neglected in general so long as $\Omega_1\tau \gg 1$. We suspect that this latter condition is not met at the shortest $\tau$, thus leading to a reduction of the coherence time. We note that this problem may be circumvented by altering the symmetry of the Solomon sequence so that all phases during consecutive cycles (repeating units of duration $\tau$, see Fig. (3a)) are inverted. Introducing the new period $\tau' = 2\tau$, it can be shown that for this modified Solomon protocol *Sl'* one gets $^{Sl'}\hat{H}_{Az}(t) = ^{Sl'}\hat{H}_{Az}(\tau'-t)$, a condition that would lead to $^{Sl'}\overline{H}_{Az}^{(k)} = 0$ for $k$ odd independently of the phase-cycle rate[22].

In summary, we have shown both through experiment and theory that one can substantially extend the rotating frame coherence lifetime of a NV center in an individual diamond nanocrystal through the Solomon protocol. The observed rotary echoes can be thought of as the rotating frame analog of Hahn's spin echoes, implying that the present scheme may find use for nano-diamond-based magnetic sensing. While this application requires that the probed field be (piece-wise) constant in the rotating frame, a broad frequency range is conceivable through proper selection of the static magnetic field. On the other hand, the long coherence lifetimes reported herein may be exploited to probe the nanocrystal dynamics when immersed in a fluid medium (as the effective microwave amplitude changes when the particle rotates[6]). In this modality, one may monitor, for example, the rotary echo signal after several Solomon phase cycles, whose amplitude—as shown in Figs. 4a—is a sensitive function of the microwave field. These and other sensing schemes are the subject of ongoing work.

We thank Dr. J. S. Hodges for his assistance during sample preparation. We acknowledge support from Research Corporation and from the National Science Foundation.

[15] J. Bylander, S. Gustavsson, F. Yan, F. Yoshihara, K. Harrabi, G. Fitch, D. G. Cory, Y. Nakamura, J-S. Tsai, W. D. Oliver, Nature Physics **7**,565(2011).

[16] J. Du, X. Rong, N. Zhao, Y. Wang, J. Yang, R. B. Liu, Nature **461**, 1265 (2009).

[17] I. Solomon, Phys. Rev. Lett. **2**, 301 (1959).

[18] G. D. Fuchs, V. V. Dobrovitski, D. M. Toyli, F. J. Heremans, D. D. Awschalom, Science **11**,1520-1522 (2009).

[19] R.R. Ernst, G. Bodenhausen, A. Wokaum, *Principles of Nuclear Magnetic Resonance in One and Two Dimensions* (Clarendon Press, Oxford, 1987).

[20] C.A Ryan, J.S. Hodges, D.G. Cory, Phys. Rev. Lett. **105**, 200402 (2010).

[21] G. de Lange, Z. H. Wang, D. Ristè, V. V. Dobrovitski, R. Hanson, Science **330**, 60 (2010).

[22] U. Haeberlen, *High Resolution NMR in Solids* (Academic Press, New York, 1976), p. 69.
12

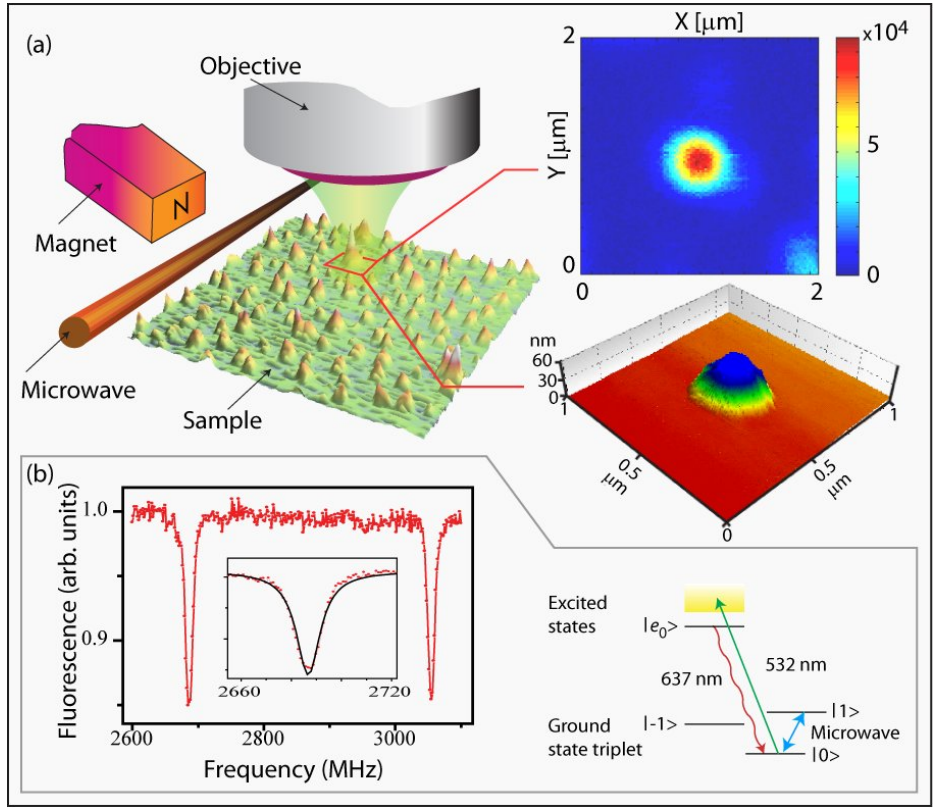

**Fig. 1:** (a) Schematics of our experimental setup. Diamond nanocrystals are uniformly distributed over a glass surface. Of these, some contain a NV center. (Lower right) AFM image of an individual 70-nm-diameter diamond particle. (Upper right) Confocal image of a NV center in nanodiamond. (b) Energy levels (right) and optically-detected magnetic resonance spectrum (left) from a NV center in a nanocrystal after a microwave frequency sweep. The figure inset zooms into one of the two possible transitions in the presence of a 6.5 mT magnetic field.

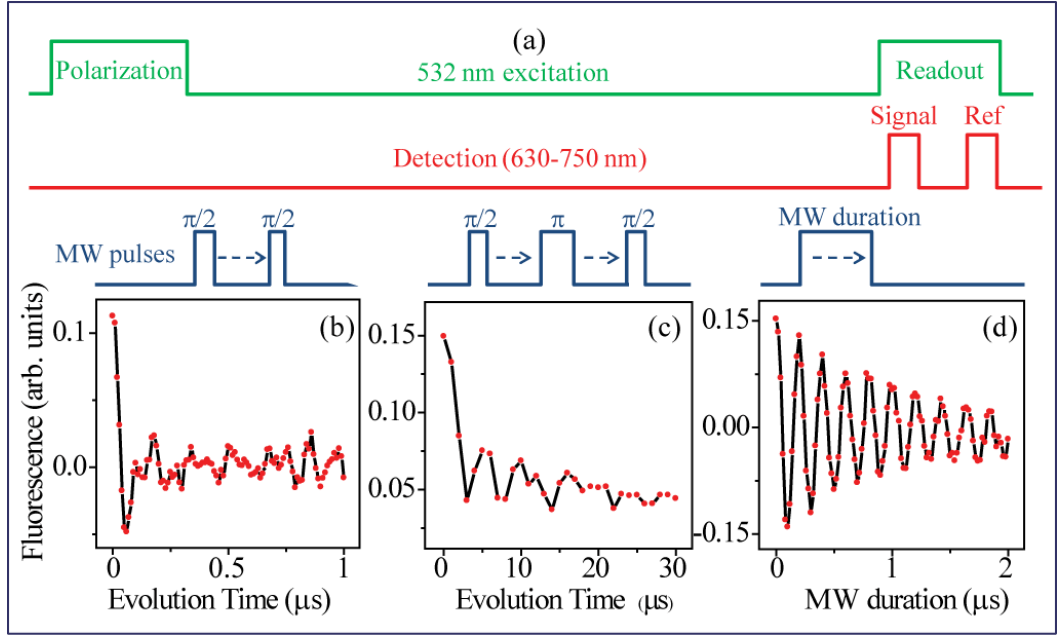

**Fig. 2:** (a) Schematics of the excitation, detection, and microwave manipulation protocols (upper, middle, and lower traces, respectively). (b)-(d) NV signal from an individual nanocrystal after the Ramsey, Hahn-echo, and Rabi sequences, respectively.

Fig. 3, Laraoui and Meriles

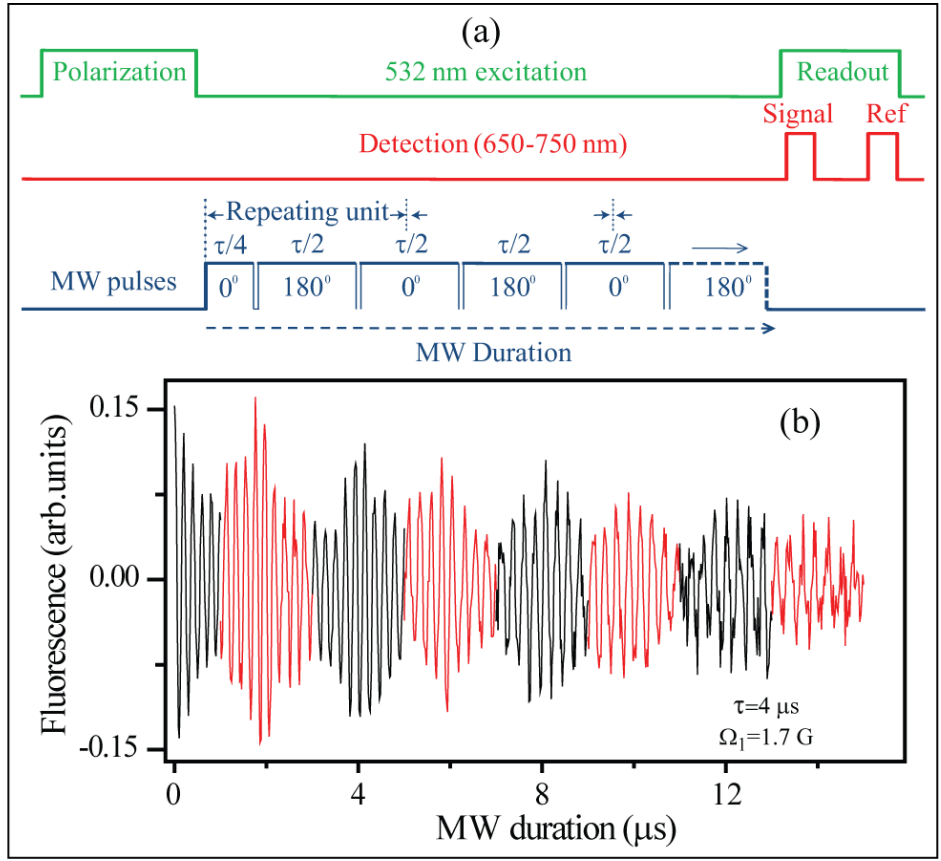

**Fig. 3:** (a) Schematics of the Solomon sequence. (b) Generation of rotary echoes from a NV center in a single diamond nanocrystal. Black and red (dark grey) segments indicate microwave excitation of opposite phase. The microwave field is 1.7 G and the phase cycle period $\tau$ is 4 μs.

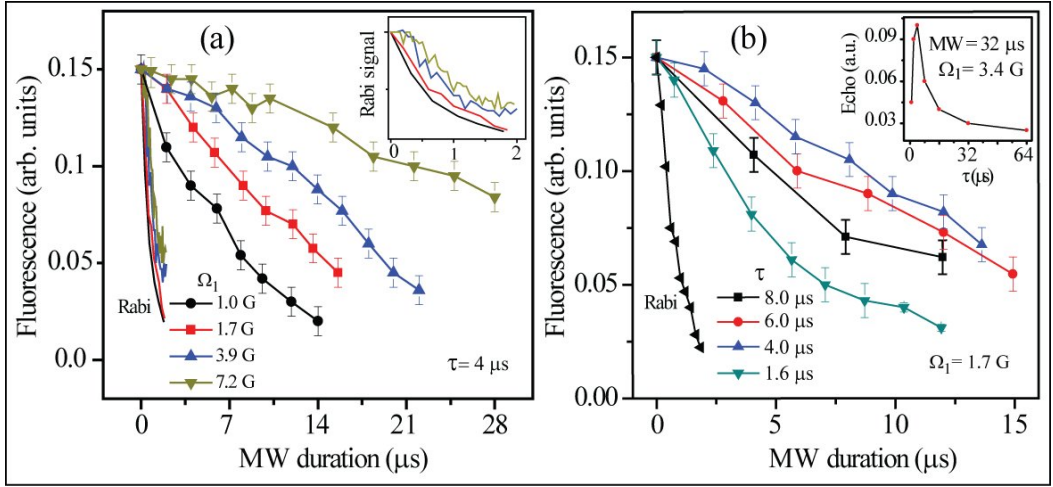

**Fig. 4:** (a) Solomon signal envelopes for various microwave field amplitudes. In all cases the phase cycle period is 4 μs. For reference, the corresponding Rabi envelopes (zoomed in the figure inset) are also reproduced using the same horizontal scale. (b) Same as in (a) but for a fixed microwave amplitude of 1.7 G and variable phase cycle period. The Rabi envelope is also included for reference. The insert shows the rotary echo amplitude as a function of $\tau$ after a fixed MW duration of 32 μs and for a microwave field of 3.4 G.